\documentclass[reprint,aps,prl,superscriptaddress]{revtex4-2}

\usepackage{graphicx}
\usepackage{amsmath}
\usepackage{siunitx}
\usepackage{bm}
\usepackage{threeparttable}
\usepackage{grffile}
\usepackage{float}

\begin{document}

\title{Robust Radiative Cooling in Functionalizable Silica Microsphere Paints}

\author{Jorge Burgos}
\affiliation{Instituto de Ciencia de Materiales de Madrid (ICMM), Consejo Superior de Investigaciones Cient\'{\i}ficas (CSIC), Calle Sor Juana In\'es de la Cruz 3, 28049 Madrid, Spain}

\author{Jos\'e Rodrigo Magana}
\affiliation{Cooling Photonics Sociedad Limitada, Barcelona, Spain}

\author{Ares Llad\'os}
\affiliation{Universitat Polit\`ecnica de Catalunya, Barcelona, Spain}

\author{Javier Pascualena Ferr\'e}
\affiliation{Universitat Polit\`ecnica de Catalunya, Barcelona, Spain}

\author{Sara N\'u\~nez-S\'anchez}
\affiliation{Instituto de Ciencia de Materiales de Madrid (ICMM), Consejo Superior de Investigaciones Cient\'{\i}ficas (CSIC), Calle Sor Juana In\'es de la Cruz 3, 28049 Madrid, Spain}

\author{Juliana Jaramillo Fern\'andez}
\affiliation{Universitat Polit\`ecnica de Catalunya, Barcelona, Spain}

\author{Cefe L\'opez}
\affiliation{Instituto de Ciencia de Materiales de Madrid (ICMM), Consejo Superior de Investigaciones Cient\'{\i}ficas (CSIC), Calle Sor Juana In\'es de la Cruz 3, 28049 Madrid, Spain}

\author{Pedro David Garc\'{\i}a}
\affiliation{Instituto de Ciencia de Materiales de Madrid (ICMM), Consejo Superior de Investigaciones Cient\'{\i}ficas (CSIC), Calle Sor Juana In\'es de la Cruz 3, 28049 Madrid, Spain}
\email{pedro.garcia@csic.es}

\begin{abstract}
Disordered coatings based on silica microspheres provide a scalable and robust platform for passive daytime radiative cooling. While particle-size optimization is often considered critical for enhancing solar scattering, the role of microsphere diameter once a coating operates in the multiple-scattering regime remains unclear. Here, we characterize the radiative cooling performance of disordered, optically thick photonic glass coatings with diameters ranging from 2 to 8~\textmu m. Despite measurable differences in microscopic scattering properties, both the spectral radiative response and the net cooling performance are robust to variations in particle diameter when the system operates deep in the diffusive regime. Outdoor thermal measurements reveal nearly identical steady-state temperature reductions across the full size range. These results indicate that radiative cooling in photonic glass coatings is governed by collective light transport, enabling microsphere size to be selected based on surface chemistry or processing constraints without compromising cooling performance.
\end{abstract}

\keywords{radiative cooling, passive daytime cooling, photonic glass, multiple light scattering, diffusive transport, silica microspheres}

\maketitle


Radiative cooling (RC) enables surfaces to dissipate heat into outer space through the atmospheric transparency window (ATW, 8--13~$\mu$m), a spectral region where atmospheric absorption is minimal. For effective below ambient daytime cooling, materials must combine solar reflectance above $90\%$ (0.3--2.5~$\mu$m) with strong thermal emission in the ATW range, such that radiative losses to the cold sky outweigh conductive, solar, and atmospheric heat gains~\cite{Hossain,Zhao,Liu,Bijarniya}. Silica (SiO$_2$) microspheres meet both requirements: their wide bandgap minimizes solar absorption, while Si--O vibrational modes provide strong mid-infrared emission in the ATW~\cite{Raman,Zhai,Atiganyanun2018}. In disordered packings of these spheres, multiple Mie scattering enhances solar reflectance while the resulting porous structure boosts thermal emissivity, enabling the performance levels summarized in Table~\ref{tab:SiO2_review}. Light in such coatings undergoes many scattering events before exiting, a regime characterized by a transport mean free path $l_t$~\cite{Atiganyanun2018,Nanophotonics2025}. When coatings are much thicker than $l_t$, transport becomes diffusive and collective, even resonant if their diameter polidispersity is low~\cite{Garcia2007}---yet whether microsphere diameter matters for radiative cooling in this limit remains unexplored. Prior work optimized particle size for single-scattering efficiency, but it is not obvious that such optimization remains relevant once diffusive transport dominates.

\begin{figure}[t!]
  \includegraphics[width=\columnwidth]{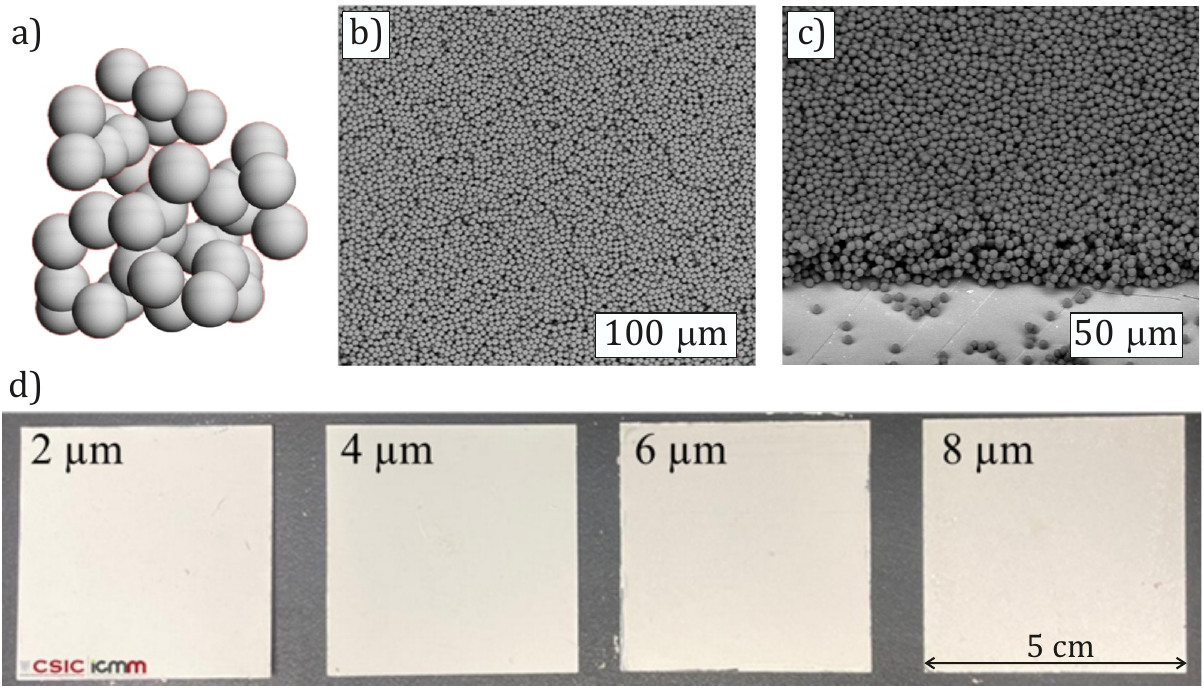}
\caption{\textbf{Morphology and structure of photonic glass coatings.}
(a)~Schematic representation of randomly packed silica microspheres forming the disordered photonic medium.
(b)~Top-view SEM image of the 4~\textmu m coating showing uniform particle distribution. 
(c)~Cross-sectional SEM image revealing the approximately 25~\textmu m thick coating on the substrate 
(d)~Optical images of all four coatings (2, 4, 6, 8~\textmu m) showing high reflectance and uniformity across the 5~$\times$~5~cm coatings.}
  \label{fig1}
\end{figure}

\begin{table*}[t]
\centering
\caption{Summary of radiative cooling performance for different SiO$_2$-based material configurations reported in the literature.}
\label{tab:SiO2_review}
\resizebox{\textwidth}{!}{%
\begin{threeparttable}
\begin{tabular}{c c c c c c c}
\hline
\textbf{Configuration} &
\textbf{Particle size} &
\textbf{Thickness} &
\textbf{Temperature drop} &
\textbf{Emissivity} &
\textbf{Cooling power} &
\textbf{Ref.} \\
 &
($\mu$m) &
($\mu$m) &
($^\circ$C) &
 &
(W/m$^2$) &
 \\
\hline
Randomly distributed in TPX polymer matrix \tnote{a} &
$\sim$4 &
50 &
-- &
$>0.93$ &
93 &
Zhai \textit{et al.} \cite{Zhai}\\
\hline
Randomly packed &
2 &
700 &
12 &
$>0.94$ &
-- &
Atiganyanun \textit{et al.} \cite{Atiganyanun2018} \\
\hline
Raspberry-like hollow spheres embedded in PDMS \tnote{b} &
1 &
2000 &
9.7 &
$>0.85$ &
114 &
Park \textit{et al.} \cite{Park2023}\\
\hline
Ordered self-assembled monolayer \tnote{c} &
8 &
8 &
19 &
0.98 &
160--350 &
Jaramillo \textit{et al.} \cite{Jaramillo2019}\\
\hline
Densely packed TiO$_2$ (top) and SiO$_2$ (bottom) &
$<0.3$ &
-- &
8 &
0.90 &
-- &
Bao \textit{et al.} \cite{Bao2017}\\
\hline
SiO$_2$ randomly distributed in PMMA matrix &
8 &
1400 &
12 &
$\sim$0.89 &
-- &
Tian \textit{et al.} \cite{Tian2025}\\
\hline
\end{tabular}
\begin{tablenotes}
\footnotesize
\item[a] Average cooling power measured around noon at $\Delta T = 0$. Below ambient cooling measurements.
\item[b] Theoretically calculated net cooling power at $\Delta T = 0$.
\item[c] Measured onto silicon substrates and at $\Delta T$ from $25$ up to 50 $^\circ$C. Showing a cooling power of 70 W/m$^2$ at $\Delta T = 0$.
\end{tablenotes}
\end{threeparttable}
}
\end{table*}

\begin{figure*}
  \includegraphics[width=\textwidth]{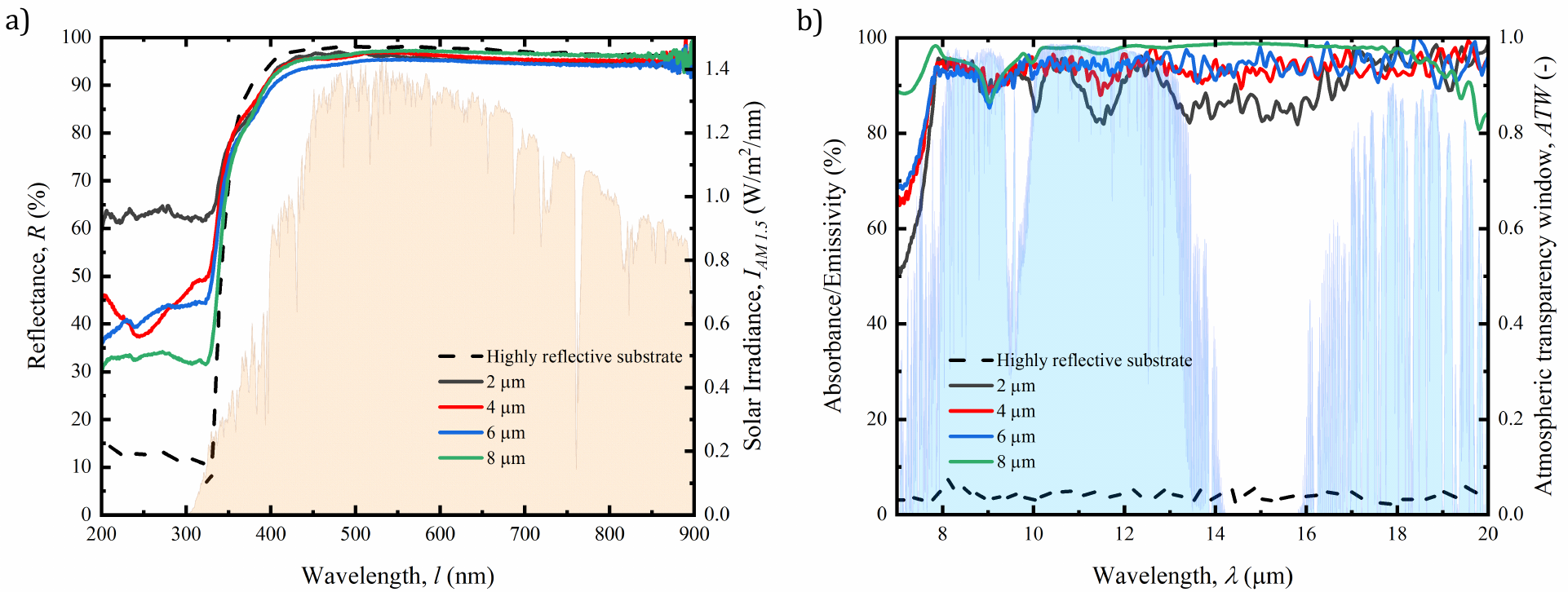}
\caption{\textbf{Optical characterization of photonic glass coatings.}
(a)~UV-Vis reflectance spectra (200--900~nm) for all coatings, overlaid with the AM1.5 solar irradiance spectrum. All coatings exhibit reflectance above 93\%, slightly lower than the bare substrate (97.1\%) but sufficient to minimize solar absorption.
(b)~Mid-infrared emissivity spectra (8--20~\textmu m) with the atmospheric transparency window highlighted. The silica coatings achieve emissivity of 0.91--0.95 in the 8--13~\textmu m atmospheric window, in contrast to the bare substrate ($\varepsilon \approx 0.02$), enabling efficient thermal radiation to outer space.}
  \label{fig2}
\end{figure*}

Here, we address this question by characterizing photonic glass coatings with diameters ranging from 2 to 8~$\mu$m. Combining optical spectroscopy, coherent backscattering measurements, and outdoor thermal characterization, we show that both spectral response and net cooling performance are largely insensitive to variations in particle diameter in the diffusive regime. These results identify diameter invariance as a defining feature of radiative cooling in diffusive coatings, enabling microsphere size to be selected based on surface chemistry or processing constraints---particularly relevant since the micrometer range offers versatile colloidal synthesis and functionalization routes~\cite{Ghimire2021} and lifting strong constraints relating to colloidal assembly. Figure~\ref{fig1} illustrates the morphology of the fabricated photonic glass coatings~\cite{Garcia2007}. Coatings with nominal microsphere diameters of 2, 4, 6, and 8~\textmu m were deposited onto substrates with solar reflectance above $95\%$ [Figure~\ref{fig2}(a)]. Scanning electron microscopy reveals uniform coatings approximately 25~\textmu m thick with randomly arranged particles, while optical images confirm complete coverage and high reflectance across all coatings (5~$\times$~5~cm$^2$ area).

The spectral radiative properties relevant to cooling performance of the photonic glass coatings are shown in Figure~\ref{fig2}. Solar reflectance spectra [Figure~\ref{fig2}(a)] show that all coatings exhibit reflectance above 93\% across the 200--900~nm range. The spectrally weighted reflectance, calculated over the measured spectral range using the AM1.5 solar spectrum, ranges from 93.6\% to 95.3\%---slightly below the bare substrate value (97.1\%) due to diffuse scattering introduced by the disordered microsphere layer, yet sufficient to minimize solar absorption. Mid-infrared emissivity spectra [Figure~\ref{fig2}(b)] show that the coatings achieve atmospheric-window emissivity (8--13~\textmu m) between 0.91 and 0.95, in stark contrast to the bare substrate (0.02). This strong infrared emission arises from the intrinsic vibrational modes of SiO$_2$ and the porous structure of the packed microsphere layer, which together produce broadband infrared emission even with a coating thickness of only 25~$\mu$m. The high emissivity within the atmospheric transparency window enables efficient radiative heat transfer to the sky.

\begin{figure*}[t!]
  \includegraphics[width=\textwidth]{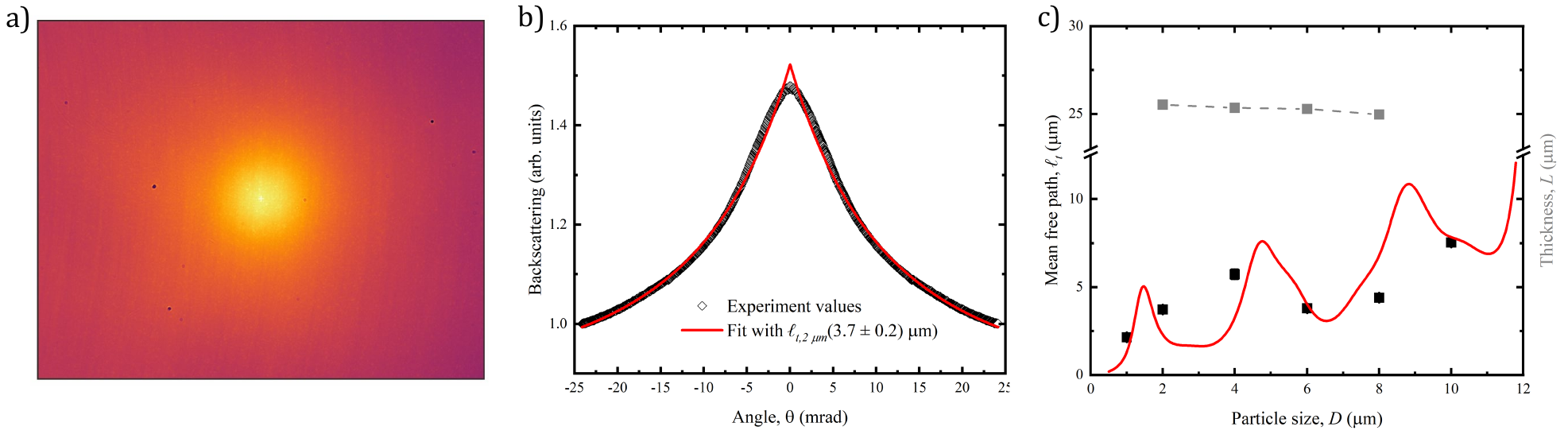}
\caption{\textbf{Light transport in disordered photonic coatings.}
(a)~Coherent backscattering cone measured from a photonic glass coating, showing multiple scattering and interference-enhanced return intensity around exact backscattering.
(b)~Coherent-backscattering profile (symbols) with fit (solid line) used to extract the transport mean free path $l_t$; the fit yields $l_t = (3.7 \pm 0.2)$~\textmu m for the 2~\textmu m coating shown.
(c)~Transport mean free path experimental values $l_t$ (black dots) as a function of microsphere diameter $D$, compared with calculated Mie Theory $l_t$ values (red curve), together with the coating thickness $L \approx 25$~\textmu m (gray horizontal line). Confirming operation deep in the multiple-scattering regime ($L \gg l_t$) for all coatings.}
  \label{fig3}
\end{figure*}

To quantify the light transport regime and confirm operation in the diffusive limit, we performed coherent backscattering measurements on all coatings. Coherent backscattering is a manifestation of weak localization in disordered media~\cite{Wolf1985,VanAlbada1985}, where constructive interference between time-reversed scattering paths produces an enhanced return signal precisely in the backscattering direction. The width of the coherent-backscattering  cone provides a direct measure of the transport mean free path $l_t$~\cite{Akkermans1986}, which characterizes the distance over which photons lose memory of their initial propagation direction and governs light diffusion in optically thick random media. Figure~\ref{fig3}(a,b) shows a representative coherent-backscattering measurement for the 2~\textmu m particle size coating. The characteristic cone-shaped enhancement around zero detection angle is clearly visible, indicating strong multiple scattering. By fitting the angular profile to diffusion theory, we extract $l_{t,2\mu m} = (3.7 \pm 0.2)$~\textmu m. Repeating this analysis for all coatings yields the transport mean free paths shown in Figure~\ref{fig3}(c) (black dots). Crucially, all coatings satisfy $L \gg l_t$, as shown in Figure~\ref{fig3}(c) where the coating thickness $L \approx 25$~\textmu m (gray dashed line) exceeds the transport length by factors ranging from approximately 3.5 (for the 8~\textmu m particles) to approximately 7 (for the 2~\textmu m particles). For comparison, the red curve in Figure~\ref{fig3}(c) shows the transport mean free path predicted by single-sphere Mie scattering theory (Supplementary information: Coherent backscattering). The experimental data (solid black squares in Figure~\ref{fig3}(c)) are not fully reproduced by the single-particle Mie prediction. This is, however, expected: at the high packing fraction of these coatings, light transport is dominated by near-field coupling and by the interaction of Mie resonances between neighboring spheres, together with the structural correlations captured by the structure factor~\cite{Aubry2017}---collective, dependent-scattering effects that a single-particle treatment can only approximate. The polydispersity of the commercial microspheres, already accounted for in the model (Supplementary Information), is a comparatively minor contribution. Here, we confirm that all coatings operate in the diffusive transport regime, where light propagation is governed by collective multiple scattering rather than by individual particle resonances.

Despite the measurable variation in transport mean free path shown in Figure~\ref{fig3}(c)---ranging from $l_t \approx 4$~\textmu m for the smallest particles to $l_t \approx 7$~\textmu m for the largest---the spectral optical properties remain remarkably consistent across all coatings. This is evident in Figure~\ref{fig2}, where both solar reflectance calculated across the measured spectral range (93.6--95.3\%) and atmospheric window emissivity (0.91--0.95) vary by less than approximately 3\% across the full diameter range investigated. This spectral diameter invariance in the diffusive regime is a central finding of our work, indicating that once the coating operates with $L \gg l_t$, radiative cooling performance becomes insensitive to the exact microsphere size. This finding is in contrast with previous theoretical results ~\cite{Whitworth2021} where the thermal emission and cooling response of periodic monolayers of identical particle sizes exhibit a pronounced diameter dependence. This effect was attributed to geometry-specific resonant effects happening only in the thin-layer limit. We next examine whether this spectral invariance translates into comparable cooling performance across the coatings.

\begin{figure*}[t!]
  \includegraphics[width=\textwidth]{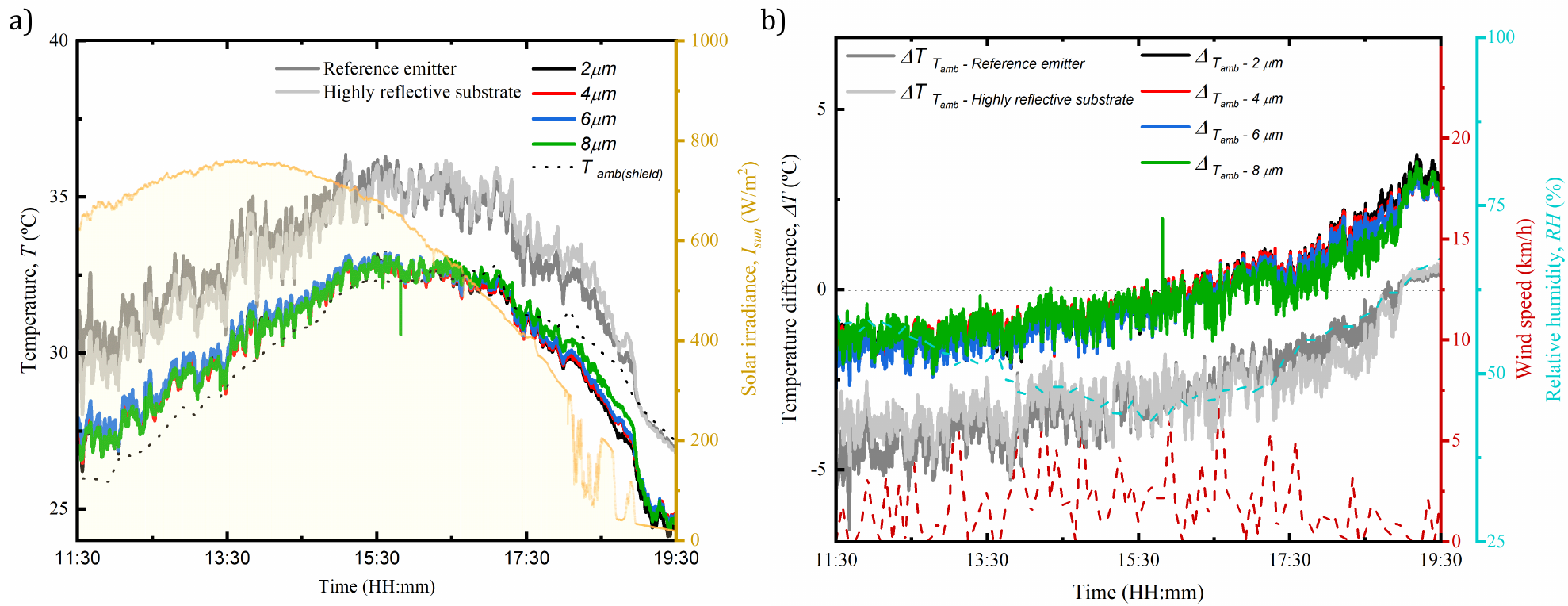}
\caption{\textbf{Thermal performance of photonic glass coatings.}
(a)~coating and reference temperatures $T$ measured over an 8-hour period from 11:30 to 19:30. Solar irradiance $I_{\text{sun}}$ (yellow, right axis) shows peak values around 800~W/m$^2$.
All photonic glass coatings (colored lines) remain at or slightly below ambient temperature ($\Delta T \approx 0$), while reference coatings (gray) heat up significantly above ambient ($\Delta T > 5~^\circ$C).
(b)~Temperature difference evolution showing near-zero $\Delta T$ for coatings throughout the day, with wind speed (gray, right axis) and relative humidity (cyan, right axis).
The coatings successfully maintain near-ambient temperatures even under solar irradiation, confirming effective passive cooling.}
  \label{thermal performance}
\end{figure*}

To assess the macroscopic cooling behavior under operating conditions, coatings and ambient temperatures were continuously monitored, in addition to humidity, wind speed, solar radiation and location. As shown in Figure~\ref{thermal performance} (a), all photonic glass coatings remain at temperatures very close to ambient even under high solar irradiance ($\approx$~800~W$\cdot$m$^{-2}$). Around midday (13{:}30), the temperature difference between the coatings and the ambient air approaches zero, indicating an effective reflection of solar heating at peak irradiance. In contrast, the highly reflective reference substrate---characterized by its high reflectance and low IR emissivity---and the benchmark radiative-cooling emitter ~\cite{Huang2022} both remain consistently hotter than the ambient temperature during the same period. Cooling below ambient is observed only after 17{:}30, when solar irradiance drops below 200~W$\cdot$m$^{-2}$. The photonic glass coatings reach a maximum temperature reduction of 4.8~$^{\circ}$C relative to the reference surfaces in the late afternoon. After sunset, the photonic glass coatings cool up to 3.7~$^{\circ}$C below ambient, whereas the references exhibit only minor cooling (approximately 0.5~$^{\circ}$C). During the experiments, wind speeds ranged from 0 to 10~km$\cdot$h$^{-1}$ and relative humidity from 30\% to 60\%. Although convective heat exchange with the air reduces the net radiative cooling effect, the coatings consistently outperform the references.

\begin{figure*}[t!]
  \includegraphics[width=\textwidth]{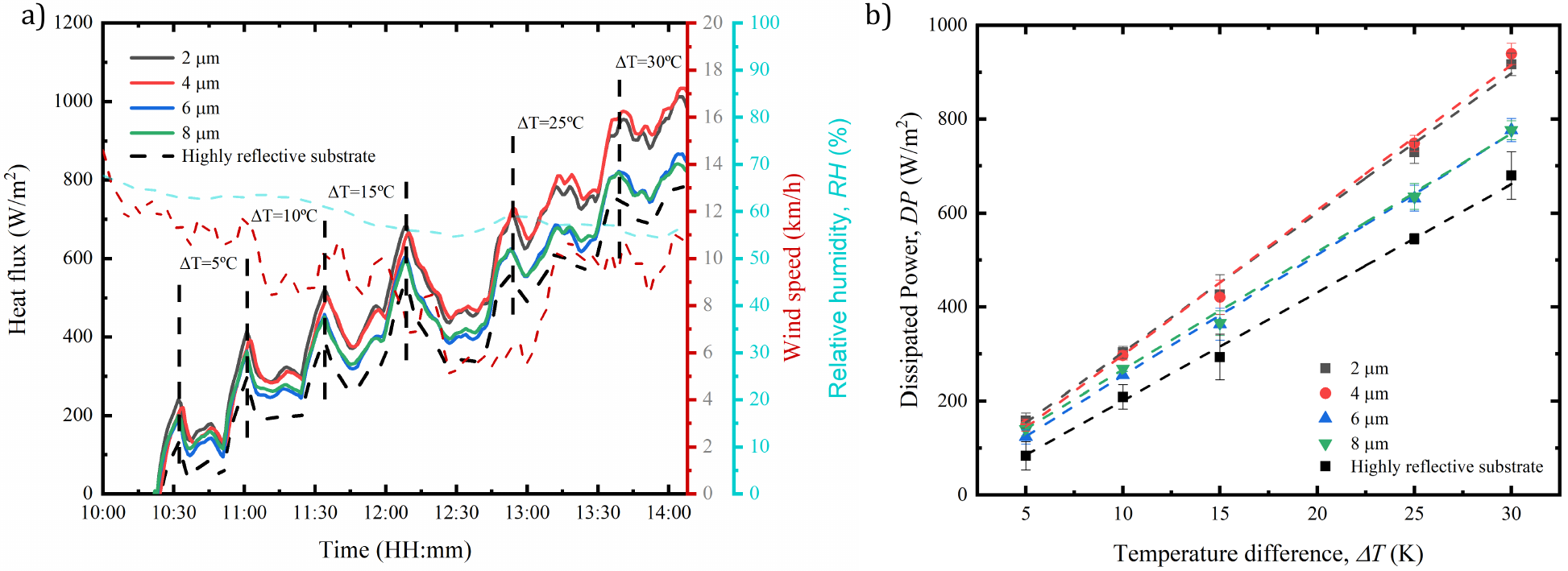}
  \caption{\textbf{Dissipated power measurements.}
  (a)~Heat flux applied to the resistive heater at steps of $\Delta T = 5, 10, 15, 20, 25, 30~^\circ$C over a 4-hour period for the four coatings (colored lines). Stabilization was reached after each temperature step. Measurements were performed under monitored conditions of wind speed (gray, left axis) and relative humidity (blue, left axis).
  (b)~Dissipated power delivered at each temperature step. Colored dots correspond to the average heat flux value at each applied temperature step. Linear fit extrapolated to $\Delta T = 0$ yields the intercept value.}
  \label{heater}
\end{figure*}

To complement the temperature-based measurements, the cooling power of each coating was evaluated directly by monitoring the dissipated power (\textit{DP}) or net cooling power capacity ($P_{\mathrm{net}}$) under outdoor conditions (see Methods – Thermal characterization: dissipated power) ~\cite{Liu,Bijarniya}. The analysis is based on the energy balance described in Equation~\ref{eq:energy_balance}:
\begin{equation}
P_{\mathrm{net}}(T_r, T_a) = P_r(T_r)-P_a(T_a)-P_{\mathrm{sun}}-P_{\mathrm{nonrad}}(T_r, T_a)
\label{eq:energy_balance}
\end{equation}
where $P_r$ is the thermal radiation emitted by the photonic glass coating, $P_a$ is the absorbed atmospheric radiation, $P_{\mathrm{sun}}$ is the absorbed solar power, and $P_{\mathrm{nonrad}}$ accounts for non-radiative heat transfer processes such as convection and conduction, with $T_r$ and $T_a$ as the radiator and ambient temperatures, respectively. The electrical power supplied to a resistive heater was measured while sequentially imposing controlled temperature differences with respect to ambient ($\Delta T = 5$--30~$^{\circ}$C), as illustrated in Figure~\ref{heater}(a). At each step, the system reached thermal equilibrium, ensuring that the supplied electrical power balanced the total heat loss from the coating. The resulting linear dependence of \textit{DP} on \(\Delta T\), shown in Figure~\ref{heater}(b), reflects the combined contribution of all the heat transfer mechanisms involved. Extrapolation to \(\Delta T = 0\) yields the baseline dissipated power under solar irradiation (daylight intercept values reported in Table~\ref{tab:results}, with an average uncertainty of \(\pm 0.3~\mathrm{W \cdot m^{-2}}\)).

It is important to note that, during the imposed temperature steps, both non-radiative heat transfer and radiative exchange with the atmosphere remain active and therefore contribute to the slope of the \textit{DP}--$\Delta T$ linear fit. In theory these temperature-dependent terms should vanish in the extrapolated limit $\Delta T = 0$, where the conditions $h_{c}(T_r - T_a) = 0$ (convection and conduction) and $\sigma\,(T_r^{4} - T_a^{4}) = 0$ (radiative exchange with the opaque atmosphere) are fulfilled. However, the real atmosphere is not a blackbody slab at ambient air temperature. It is vertically stratified, spectrally selective, humidity-dependent, and more critically its effective emitting temperature generally differs from near-surface ambient air temperature. As a consequence, the daylight intercept retains exclusively the radiative terms and the absorbed solar radiation.

We complemented these measurements with nighttime experiments under clear-sky conditions (20{:}00--22{:}00), with no solar irradiance and the heater set to maintain $\Delta T = 0$. Under these conditions, the measured dissipated power (as shown in Figure~S3 of the Supplementary Information) reflects primarily radiative exchange with the sky (atmosphere and space), yielding a direct determination of the radiative cooling power ($P_{\mathrm{COOL,night}}$), as shown in Table~\ref{tab:results}. The \textit{difference} between $P_{\mathrm{COOL,night}}$ and the daylight intercept approximates the net energy balance under solar exposure, assuming negligible differences in relative humidity; positive daylight intercept values indicating the potential to achieve sub-ambient cooling during daytime operation. However, deviations from the ideal energy balance arise from parasitic heat losses through the test box, convective effects, and local variations in outdoor conditions (e.g., wind, humidity, and cloud cover). These contributions result in a standard deviation of approximately $\pm 20~\mathrm{W/m^2}$, corresponding to up to $\sim$20 \%  of the expected net cooling power. Across the four coatings, these night-time cooling powers (44--61~W/m$^2$) show no systematic dependence on microsphere diameter, consistent with the spectral invariance reported above.

As an initial approximation, and assuming that non-radiative heat-transfer mechanisms are negligible, the effective solar absorption of the coatings can be estimated from the thermal measurements (reported as Estimated solar absorption in Table~II). The thermally derived values were subsequently normalized to standard conditions of 800~W\,m$^{-2}$ and 25~$^{\circ}$C in order to enable a meaningful comparison with the solar absorption measured in the 200--900~nm wavelength range. The (measured) solar absorption was calculated as $a_{\mathrm{total}} = 1 - R_{\mathrm{material}}$, where $R_{\mathrm{material}}$ is the integrated reflectance obtained from Equation~2. This normalization procedure is consistent with the linearized heat-balance treatment described in the Supplementary Information (Data Normalization for Dissipated Power Measurements), where the measured dissipated power is interpreted through an effective loss coefficient that accounts for the combined convective and radiative contributions under the outdoor test conditions. By rescaling the thermal data to a common reference state, the thermally estimated and spectroscopically measured solar absorptions agree in magnitude, both lying in the 5--8\% range. The match is closest for the 4 and 6~\textmu m coatings, while the thermal estimate exceeds the measured value by about 2 percentage points for the 2 and 8~\textmu m coatings---reflecting the approximate nature of the thermal estimate, which neglects non-radiative transfer. This broad consistency supports the reliability of the thermal measurements and the observed temperature-drop performance.

\begin{table}[b]
\caption{Thermal characterization results for photonic glass coatings.}
\label{tab:results}
\centering
\begin{tabular}{lcccc}
\hline
coating & 2~\textmu m & 4~\textmu m & 6~\textmu m & 8~\textmu m \\
\hline
Intercept daylight (W/m$^2$) & $-0.1$ & $0.5$ & $0.2$ & $-0.6$ \\
$P_{\text{cool,night}}$ (W/m$^2$) & 60.8 & 46.2 & 44.1 & 54.8 \\
Difference (W/m$^2$) & 61.0 & 45.7 & 43.9 & 55.4 \\
Estimated solar absorption (\%) & 7.6 & 5.7 & 5.5 & 6.9 \\
Measured absorption (\%) & 5.4 & 5.2 & 6.4 & 4.8 \\
\hline
\end{tabular}
\end{table}

The experimental results presented here show that radiative cooling performance of photonic glass coatings remains largely stable across a broad range of particle diameters. This behavior can be understood as a direct consequence of operating in the diffusive transport regime, where the coating thickness greatly exceeds the transport mean free path ($L \gg l_t$). In this limit, light undergoes many scattering events (about 7 $\sim L / l_t$) before exiting the material, and the macroscopic optical response is governed by collective transport properties rather than the scattering characteristics of individual particles. As a result, variations in microsphere diameter primarily affect microscopic parameters such as $l_t$, while leaving the overall emissivity in the ATW (between 8 to 13~$\mu$m)  and solar reflectance (measured visible wavelength range) essentially unchanged.
This interpretation is consistent with the observed invariance of the measured spectra, passive temperature reduction, and net cooling power, and aligns with previous studies of radiative cooling in optically thick disordered media ~\cite{Atiganyanun2018}. Importantly, this collective transport regime relaxes the dependence of radiative cooling on precise particle-size control.

In conclusion, we report passive radiative cooling using functionalizable photonic glass coatings, achieving sub-ambient temperature reductions under direct sunlight and net cooling powers lying within the same order of magnitude as those reported for state-of-the-art radiative coolers. By systematically varying the microsphere diameter and combining spectral, thermal, and power-based measurements across particle sizes spanning the visible and infrared scattering regimes, we find the cooling performance to be equal within experimental error. This robustness, rooted in collective light transport in optically thick coatings, relaxes fabrication constraints and, crucially, decouples radiative cooling performance from precise control over particle size. As a result, radiative cooling can be integrated with the full toolbox of colloidal chemistry, enabling functionalization, surface modification, and chemical tuning without compromising thermal performance, and opening pathways toward scalable, multifunctional cooling materials.

\section*{Methods}

\textbf{Photonic glass coating fabrication.} Tested coatings were prepared by depositing polydisperse suspensions of commercial silica microspheres (Microparticles GmbH) with nominal diameters of 2, 4, 6, and 8~\textmu m onto highly reflective substrates. Coatings of 5~$\times$~5~cm were fabricated using dispersions of 15~mL with concentrations of 0.4~wt\% for coatings of 2 and 8~\textmu m in diameter, and 0.5~wt\% for 4 and 6~\textmu m in diameter. These concentrations were identified as necessary to maintain a thickness of approximately 25~\textmu m across all four particle sizes, with a standard deviation of $\pm\,0.23~\mu\text{m}$. The substrates were placed in individual boxes of the same area and a controlled environment where the dispersion was allowed to settle naturally through a gravity-assisted process. Drying at 45~$^\circ$C and 30\% relative humidity yielded mechanically stable layers with excellent uniformity and a thickness of approximately 25~\textmu m.

\textbf{Optical spectroscopy.} UV-Vis reflectance was measured using a UV-Vis spectrometer (Cary 4000, Agilent Technologies) equipped with an integrating sphere accessory (Internal DRA 900). A barium-sulfate-based diffuse reflectance standard was used as reference. The spectrally weighted solar reflectance was calculated as:
\begin{equation}
R(\%) = \frac{\int_{200}^{900} I_{\text{sun}}(\lambda)\, R_{\text{material}}(\lambda)\, d\lambda}{\int_{200}^{900} I_{\text{sun}}(\lambda)\, d\lambda},
\end{equation}
where $I_{\text{sun}}$ is the AM1.5 solar irradiance spectrum and $R_{\text{material}}$ is the measured reflectance.

Mid-infrared emissivity was measured using FTIR spectroscopy (Vertex 80, Bruker) coupled with a gold-coated integrating sphere accessory (A562, Bruker), with a gold mirror as reference. Since the coatings are optically opaque, emissivity equals absorptance $A = 1 - R$ by Kirchhoff's law. The atmospheric window emissivity was calculated as:
\begin{equation}
\varepsilon = \frac{\int_{8}^{13} bb(\lambda,T)\, A_{\text{material}}(\lambda)\, d\lambda}{\int_{8}^{13} bb(\lambda,T)\, d\lambda},
\end{equation}
where $bb(\lambda,T)$ is the blackbody spectral radiance at temperature $T$ and $A_{\text{material}} = 1 - R_{\text{material}}$.

\textbf{Coherent backscattering measurements.} Coherent-backscattering measurements were performed using a He-Ne laser source. The beam was spatially filtered to select the fundamental TEM$_{00}$ mode, then directed through a beamsplitter onto the coating. A quarter-wave plate positioned between the beamsplitter and the coating created circular polarization for the incident light. Backscattered light passed through the same quarter-wave plate and beamsplitter before reaching a CCD camera. A linear polarizer placed before the camera filtered single-scattered photons (which have flipped helicity) while transmitting multiply-scattered light (which randomizes polarization). The angular profile of the coherent backscattering cone was recorded and fitted to diffusion theory~\cite{Akkermans1986} to extract the transport mean free path $l_t$.

\textbf{Thermal characterization: temperature drop.} Temperature-drop measurements were performed under clear-sky outdoor conditions in Barcelona (Spain, October 2025). Photonic glass coatings were mounted on a wooden platform covered with aluminized Mylar and oriented south, with sufficient spacing to avoid mutual shading (as shown in Figure~S1 of the Supplementary Information). The platform was fixed to a PVC tube anchored to a brick base, elevating the coatings 1.5~m above a flat concrete floor and at least 10~m away from any vertical structure to eliminate shading effects. Each coating was attached to its holder using double-sided adhesive tape and mounted on an expanded polystyrene support covered with a highly reflective, low-emissivity polymer film to minimize parasitic heat gains through the support structure. No side walls, additional thermal insulation, or convection-shield layers were used, ensuring an unobstructed view of the sky and allowing natural convective and radiative exchange with the surrounding environment. The configuration accommodated up to six coatings simultaneously, including the disordered silica sphere photonic glass coatings fabricated in this study, a highly reflective substrate used as a reference near-perfect solar reflector, and a known radiative cooling standard for benchmarking~\cite{Huang2022}.

Temperature measurements were performed using calibrated NTC thermistors; temperature sensors were Class~A, 4-wire PT100 probes (RSPro). Sensors monitoring coating temperature were bonded with conductive silver paint to the center of the backside of each coating. An additional sensor, housed in a Th-Friedrichs type~2033 radiation shield, measured ambient air temperature at the same height as the coatings and adjacent to a weather station (7002571 PC Weather Station, Bresser), which recorded relative humidity, wind speed, and other atmospheric parameters at 5-min intervals. Global horizontal irradiance was measured with a Class~B pyranometer (CMP6, Kipp~\&~Zonen) placed within 0.5~m of the setup. All temperatures and irradiance values were logged using a data acquisition system (DAQ970A, Keysight) with a sampling interval of 10~s.

\textbf{Thermal characterization: dissipated power.} Dissipated power was measured using an identical custom-built outdoor experimental system (as shown in Figure~S2 of the Supplementary Information). A resistive heating pad controlled by a PID loop implemented on a microcontroller maintained the coating surface at a fixed temperature setpoint above ambient, imposing a controlled temperature difference ($\Delta T$) between the coating and the surrounding air. The electrical power supplied to the heater was continuously recorded and represents the energy required to offset heat losses from the coating.

\section*{Associated Content}
\noindent\textbf{Supporting Information}\\
Mie-theory analysis of the transport mean free path from coherent backscattering; photographs of the outdoor temperature-drop and dissipated-power experimental setups; night-time heat-flux measurements; and the data-normalization procedure for the dissipated-power measurements (PDF).

\section*{Acknowledgments}

This work was supported by the Spanish Ministry of Science, Innovation and Universities via the national project PID2024-158832NB-C21 (PSYNC), by the HORIZON-EIC-2022 Pathfinder project ADAPTATION (Grant No.~101129661), and by the Spanish MICIU Severo Ochoa program for Centres of Excellence through Grant CEX2024-001445-S. The Table of Contents graphic was created with the assistance of a large-language model.

\onecolumngrid
\clearpage

\setcounter{figure}{0}
\setcounter{table}{0}
\setcounter{equation}{0}
\renewcommand{\thefigure}{S\arabic{figure}}
\renewcommand{\theequation}{S\arabic{equation}}
\renewcommand{\thetable}{S\arabic{table}}

\begin{center}
{\large\textbf{Supplementary Information}}\\[2pt]
\textbf{Robust Radiative Cooling in Functionalizable Silica Microsphere Paints}
\end{center}

\vspace{6pt}

\subsection*{Coherent Backscattering}

The black symbols show the transport mean free path $l_t$ obtained from coherent
backscattering (CBS) measurements for photonic glasses with different particle
diameters. The red line corresponds to the single-particle Mie prediction for
the transport cross section,
\begin{equation}
\sigma_t(d) = \sigma_s(d)\,[1-g(d)],
\end{equation}
where $\sigma_s$ and $g$ are the scattering cross section and anisotropy factor
of an isolated SiO$_2$ sphere in air. To account for the measured
polydispersity of the colloids, the Mie transport cross section is averaged over
a log-normal size distribution with 2\% relative width. This suppresses the
extremely sharp resonances present in the monodisperse Mie curves and yields a
smooth theoretical trend for $l_t(d)$.

Since the samples have a high packing fraction ($\phi \simeq 0.5$), the
independent-scatterer approximation is not sufficient. Structural correlations
and near-field coupling modify the angular scattering distribution through the
static structure factor $S(q)$, effectively enhancing large-angle scattering and
reducing $l_t$ compared to the dilute limit. A full calculation,
\begin{equation}
l_t^{-1} = \pi k^6 \int_0^{2k} F(q)\,S(q)\,q^3\,\mathrm{d}q\,\rho,
\end{equation}
where $F(q)$ is the Mie form factor, is beyond the scope of this figure. Instead,
following previous analyses of dense photonic glasses, these dependent-scattering
effects are represented by a single global multiplicative factor applied to the
polydisperse Mie prediction. The scaled curve reproduces the correct magnitude
and the increasing trend of $l_t(d)$ and illustrates that light transport in
photonic glasses cannot be described by independent scattering alone.

\subsection*{Thermal Characterisation: Temperature Drop}

\begin{figure}[H]
    \centering
    \includegraphics[width=0.5\linewidth]{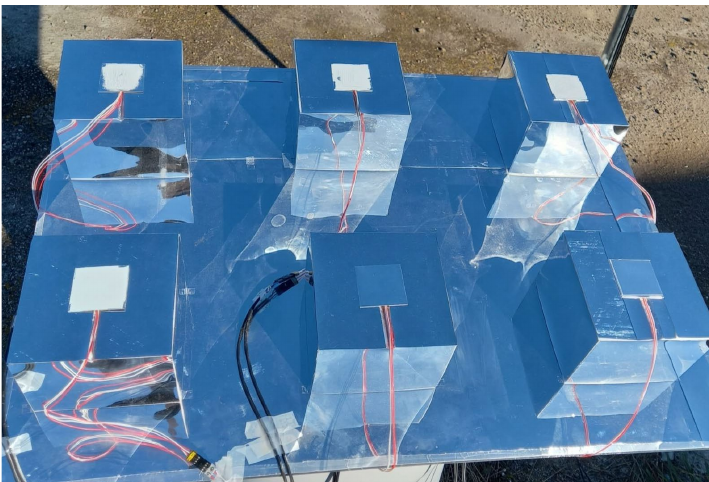}
    \caption{Experimental setup for temperature drop and power cooling performance measurements.}
    \label{fig:S1}
\end{figure}

Figure~\ref{fig:S1} shows the custom-built outdoor platform during the temperature-drop measurement. The four photonic glass coatings are mounted side by side with the highly reflective reference and the radiative-cooling benchmark, with sufficient spacing to avoid mutual shading. No side walls, thermal insulation, or convection shields are used, so that each sample exchanges heat with the clear sky and the surrounding air under realistic, unobstructed conditions, while a low-emissivity support thermally decouples it from the platform (full instrumentation in Methods). No power is supplied to the samples in this configuration, so the back-side temperatures recorded under clear skies give directly the steady-state temperature difference between each coating and the ambient air reported in Figure~4 of the main text. The same platform, equipped with a controlled resistive heater, is used for the dissipated-power measurements of Figure~\ref{fig:S2}.

\subsection*{Thermal Characterization: Dissipated Power}

\begin{figure}[H]
    \centering
    \includegraphics[width=0.5\linewidth]{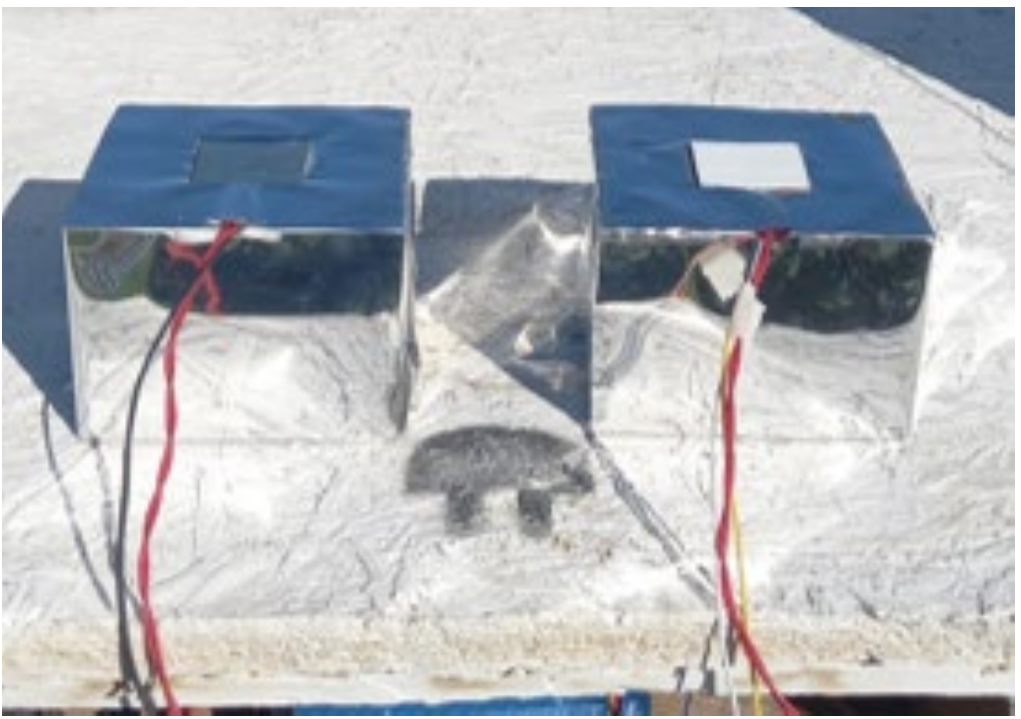}
    \caption{Dissipated power experimental setup with a resistive heating pad controlled by a PID loop implemented on a microcontroller.}
    \label{fig:S2}
\end{figure}

Figure~\ref{fig:S2} shows the same platform in the dissipated-power configuration. A resistive heating pad bonded to the back of each sample is driven by a PID loop implemented on a microcontroller, which holds the sample at a fixed temperature setpoint above the ambient air. The electrical power supplied to maintain a given over-temperature $\Delta T$ equals the net heat lost by the sample at that $\Delta T$; stepping $\Delta T$ and recording the supplied power once thermal equilibrium is reached at each step yields the dissipated-power--$\Delta T$ relation analysed through the main-text energy balance (Eq.~1).

\begin{figure}[H]
    \centering
    \includegraphics[width=0.55\linewidth]{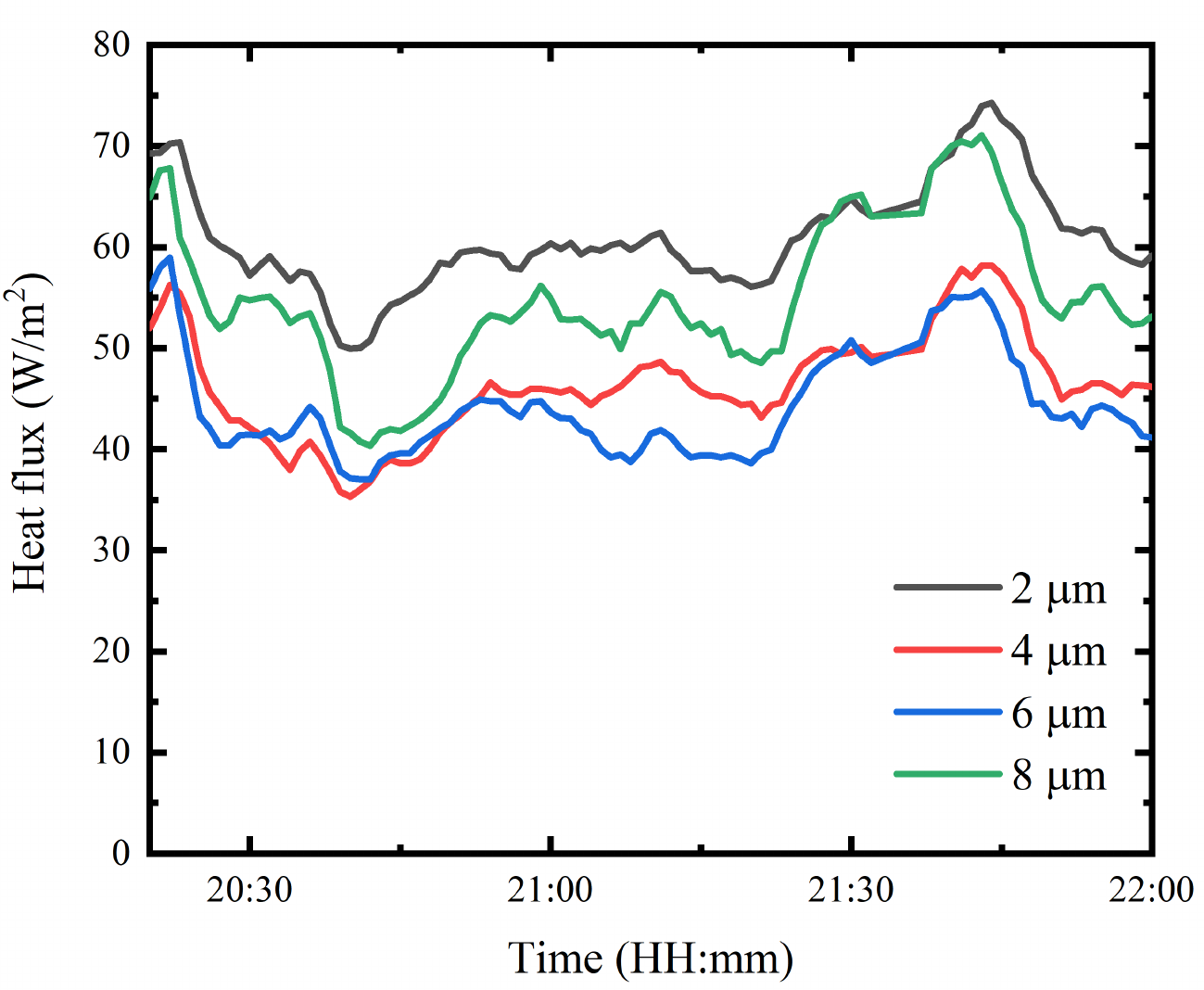}
    \caption{Nighttime heat flux measurements keeping $\Delta T = 0$ controlled by the PID loop.}
    \label{fig:S3}
\end{figure}

Figure~\ref{fig:S3} shows the corresponding night-time measurement. Under clear-sky conditions between 20{:}00 and 22{:}00, with no solar irradiance and the PID loop holding $\Delta T = 0$, convective and conductive exchange are suppressed, so that the recorded dissipated power reflects primarily the radiative exchange between each coating and the sky. This provides the direct determination of the radiative cooling power $P_{\mathrm{cool,night}}$ listed in Table~II of the main text. The recorded heat fluxes are of the same magnitude as those $P_{\mathrm{cool,night}}$ values, with the $2~\mu$m coating consistently the highest; the slow drift common to all four traces follows the gradual change in effective sky temperature and humidity over the measurement, whereas the modest, non-monotonic differences between coatings are consistent with the diameter invariance discussed in the main text.

\subsection*{Data Normalization for Dissipated Power Measurements}

Given that the measurements are dependent on external conditions (wind, ambient temperature, solar irradiance), we normalize the measurements to fixed reference conditions of ambient temperature and solar irradiance. Each record contains the dissipated power of the samples $P\,[\mathrm{W\,m^{-2}}]$, the plane-of-sample global solar irradiance $G\,[\mathrm{W\,m^{-2}}]$, the ambient air temperature $T_a$ and the surface temperature of the sample $T_s\,[\mathrm{K}]$, together with the sample's spectroscopic total solar absorptance $\alpha$ and long-wave emissivity $\varepsilon$. The over-temperature is $\Delta T = T_s - T_a$. The steady, per-area heat balance during the day is written as:
\begin{equation}
P + \alpha G = h_c\,\Delta T + q_{\mathrm{LW}},
\end{equation}
where $h_c\,[\mathrm{W\,m^{-2}\,K^{-1}}]$ is an effective convective coefficient that embodies wind and geometry, and $q_{\mathrm{LW}}\,[\mathrm{W\,m^{-2}}]$ is the net long-wave exchange with the sky and surroundings. This expression represents the linearized form of the general energy-balance relation introduced as Equation~1 in the main manuscript, in which the temperature-dependent radiative contribution is approximated to first order to facilitate the interpretation of the outdoor thermal measurements. The long-wave term for $\Delta T \lesssim 30\ \mathrm{K}$ is linearized about the mean-film temperature $T_m = (T_s + T_a)/2$, yielding
\begin{equation}
q_{\mathrm{LW}} \approx h_r\,\Delta T
\quad\text{with}\quad
h_r = 4 \varepsilon \sigma T_m^3,
\end{equation}
and $\sigma = 5.670374419 \times 10^{-8}\ \mathrm{W\,m^{-2}\,K^{-4}}$. The accuracy of the mean-film linearization is within one percent for $\Delta T \lesssim 30\ \mathrm{K}$. Substituting gives an effective loss coefficient as a sum of conduction/convection heat transfer coefficient and the radiant one.
\begin{equation}
 \frac{P + \alpha G}{\Delta T} \approx h_c + h_r,
\end{equation}
This inference step anchors the normalization in measured aerothermal conditions rather than assumed correlations.

Normalization to a common ambient temperature and irradiance proceeds by reconstructing the same balance at the chosen reference state of reference ambient temperature and solar irradiance $(T_{a,\mathrm{ref}}, G_{\mathrm{ref}})$. The target equals the measured $\Delta T$ of each record, preserving the setpoint achieved. At the reference state we update long-wave radiation to the new temperatures via
\begin{equation}
T_{s,\mathrm{ref}} = T_{a,\mathrm{ref}} + \Delta T_{\mathrm{ref}}, \qquad
T_{m,\mathrm{ref}} = \frac{T_{s,\mathrm{ref}} + T_{a,\mathrm{ref}}}{2},
\end{equation}
and
\begin{equation}
h_{r,\mathrm{ref}} = 4 \varepsilon \sigma T_{m,\mathrm{ref}}^3.
\end{equation}
We preserve the measured convection by carrying over the inferred $h_c$. The normalized dissipated power $P_{\mathrm{ref}}$ then follows directly:
\begin{equation}
P_{\mathrm{ref}} = (h_c + h_{r,\mathrm{ref}})\,\Delta T - \alpha\,G_{\mathrm{ref}}.
\end{equation}
This normalization removes day-to-day variability of sun and ambient while retaining the measured convective character of each run.

At night the short-wave term vanishes and $\Delta T \simeq 0$ eliminates convection, so the balance collapses to the radiative deficit against the sky,
\begin{equation}
P = \varepsilon \sigma \left(T_a^4 - T_{\mathrm{sky}}^4\right).
\end{equation}
The effective sky temperature $T_{\mathrm{sky}}$ is inverted from the measurement as
\begin{equation}
T_{\mathrm{sky}} = \left(T_a^4 - \frac{P}{\varepsilon \sigma}\right)^{1/4}.
\end{equation}
The quantity $\Delta T_{\mathrm{sky}} = T_a - T_{\mathrm{sky}}$ is a radiative contrast that characterizes the nocturnal environment. Assuming this contrast is the invariant descriptor to carry between ambients, we form
\begin{equation}
T_{\mathrm{sky,ref}} = T_{a,\mathrm{ref}} - \Delta T_{\mathrm{sky}}
\end{equation}
and evaluate the normalized nocturnal requirement at $(T_{a,\mathrm{ref}}, G_{\mathrm{ref}} = 0)$ and $\Delta T_{\mathrm{ref}} = 0$:
\begin{equation}
P_{\mathrm{ref,night}} = \varepsilon \sigma \left(T_{a,\mathrm{ref}}^4 - T_{\mathrm{sky,ref}}^4\right).
\end{equation}

In practice, nighttime rows are identified by small irradiance and near-zero $\Delta T$. When noise corrupts the inversion, a clear-sky surrogate $T_{\mathrm{sky}} \approx T_a - \Delta T_{\mathrm{sky,0}}$ with $\Delta T_{\mathrm{sky,0}}$ in the $10$--$20\ \mathrm{K}$ range provides a stable fallback.


\begin{thebibliography}{99}

\bibitem{Hossain}
M.~M.~Hossain and M.~Gu, Radiative cooling: Principles, progress, and potentials. \textit{Adv.\ Sci.}\ \textbf{2016}, \textit{3} (7), 1500360. DOI: 10.1002/advs.201500360.

\bibitem{Zhao}
B.~Zhao, M.~Hu, X.~Ao, N.~Chen, and G.~Pei, Radiative cooling: A review of fundamentals, materials, applications, and prospects. \textit{Appl.\ Energy} \textbf{2019}, \textit{236}, 489--513. DOI: 10.1016/j.apenergy.2018.12.018.

\bibitem{Liu}
P.~Liu \textit{et al.}, Functional radiative cooling: Basic concepts, materials, and best practices in measurements. \textit{ACS Appl.\ Electron.\ Mater.}\ \textbf{2023}, \textit{5} (12), 6395--6422. DOI: 10.1021/acsaelm.3c01023.

\bibitem{Bijarniya}
J.~P.~Bijarniya, J.~Sarkar, and P.~Maiti, Review on passive daytime radiative cooling: Fundamentals, recent researches, challenges and opportunities. \textit{Renew.\ Sustain.\ Energy Rev.}\ \textbf{2021}, \textit{143}, 110263. DOI: 10.1016/j.rser.2020.110263.

\bibitem{Raman}
A.~P.~Raman, M.~A.~Anoma, L.~Zhu, E.~Rephaeli, and S.~Fan, Passive radiative cooling below ambient air temperature under direct sunlight. \textit{Nature} \textbf{2014}, \textit{515} (7528), 540--544. DOI: 10.1038/nature13883.

\bibitem{Zhai}
Y.~Zhai, Y.~Ma, S.~N.~David, D.~Zhao, R.~Lou, G.~Tan, R.~Yang, and X.~Yin, Scalable-manufactured randomized glass--polymer hybrid metamaterial for daytime radiative cooling. \textit{Science} \textbf{2017}, \textit{355} (6329), 1062--1066. DOI: 10.1126/science.aai7899.

\bibitem{Atiganyanun2018}
S.~Atiganyanun, J.~B.~Plumley, S.~J.~Han, K.~Hsu, J.~Cytrynbaum, T.~L.~Peng, S.~M.~Han, and S.~E.~Han, Effective radiative cooling by paint-format microsphere-based photonic random media. \textit{ACS Photonics} \textbf{2018}, \textit{5} (4), 1181--1187. DOI: 10.1021/acsphotonics.7b01492.

\bibitem{Nanophotonics2025}
J.~Kang \textit{et al.}, Design strategies, manufacturing, and applications of radiative cooling technologies. \textit{Nanophotonics} \textbf{2025}, \textit{14} (15), 2355--2395. DOI: 10.1515/nanoph-2025-0159.

\bibitem{Garcia2007}
P.~D.~Garc\'{\i}a, R.~Sapienza, \'A.~Blanco, and C.~L\'opez, Photonic glass: A novel random material for light. \textit{Adv.\ Mater.}\ \textbf{2007}, \textit{19} (18), 2597--2602. DOI: 10.1002/adma.200602426.

\bibitem{Ghimire2021}
P.~P.~Ghimire and M.~Jaroniec, Renaissance of St\"ober method for synthesis of colloidal particles: New developments and opportunities. \textit{J.\ Colloid Interface Sci.}\ \textbf{2021}, \textit{584}, 838--865. DOI: 10.1016/j.jcis.2020.10.014.

\bibitem{Park2023}
C.~Park, C.~Park, S.~Park, J.~Lee, Y.~S.~Kim, and Y.~Yoo, Hybrid emitters with raspberry-like hollow SiO$_2$ spheres for passive daytime radiative cooling. \textit{Chem.\ Eng.\ J.}\ \textbf{2023}, \textit{459}, 141652. DOI: 10.1016/j.cej.2023.141652.

\bibitem{Jaramillo2019}
J.~Jaramillo-Fernandez, G.~L.~Whitworth, J.~A.~Pariente, A.~Blanco, P.~D.~Garcia, C.~Lopez, and C.~M.~Sotomayor-Torres, A self-assembled 2D thermofunctional material for radiative cooling. \textit{Small} \textbf{2019}, \textit{15} (52), 1905290. DOI: 10.1002/smll.201905290.

\bibitem{Bao2017}
H.~Bao, C.~Yan, B.~Wang, X.~Fang, C.~Y.~Zhao, and X.~Ruan, Double-layer nanoparticle-based coatings for efficient terrestrial radiative cooling. \textit{Sol.\ Energy Mater.\ Sol.\ Cells} \textbf{2017}, \textit{168}, 78--84. DOI: 10.1016/j.solmat.2017.04.020.

\bibitem{Tian2025}
X.~Tian, H.~Wang, Y.~Lu, M.~Wang, J.~Wang, H.~Liu, W.~Zhou, G.~Zhao, J.~Gao, F.~Sun, X.~Meng, and Z.~Qu, Auto-deposited microparticle composite coating for low-cost and efficient daytime radiative cooling. \textit{ACS Appl.\ Mater.\ Interfaces} \textbf{2025}, \textit{17} (5), 8274--8284. DOI: 10.1021/acsami.4c18499.

\bibitem{Wolf1985}
P.-E.~Wolf and G.~Maret, Weak localization and coherent backscattering of photons in disordered media. \textit{Phys.\ Rev.\ Lett.}\ \textbf{1985}, \textit{55} (24), 2696--2699. DOI: 10.1103/PhysRevLett.55.2696.

\bibitem{VanAlbada1985}
M.~P.~van Albada and A.~Lagendijk, Observation of weak localization of light in a random medium. \textit{Phys.\ Rev.\ Lett.}\ \textbf{1985}, \textit{55} (24), 2692--2695. DOI: 10.1103/PhysRevLett.55.2692.

\bibitem{Akkermans1986}
E.~Akkermans, P.~E.~Wolf, and R.~Maynard, Coherent backscattering of light by disordered media: Analysis of the peak line shape. \textit{Phys.\ Rev.\ Lett.}\ \textbf{1986}, \textit{56} (14), 1471--1474. DOI: 10.1103/PhysRevLett.56.1471.

\bibitem{Aubry2017}
G.~J.~Aubry, L.~Schertel, M.~Chen, H.~Weyer, C.~M.~Aegerter, S.~Polarz, H.~C\"olfen, and G.~Maret, Resonant transport and near-field effects in photonic glasses. \textit{Phys.\ Rev.\ A} \textbf{2017}, \textit{96} (4), 043871. DOI: 10.1103/PhysRevA.96.043871.

\bibitem{Whitworth2021}
G.~L.~Whitworth, J.~Jaramillo-Fernandez, J.~A.~Pariente, P.~D.~Garcia, A.~Blanco, C.~Lopez, and C.~M.~Sotomayor-Torres, Simulations of micro-sphere/shell 2D silica photonic crystals for radiative cooling. \textit{Opt.\ Express} \textbf{2021}, \textit{29} (11), 16857--16866. DOI: 10.1364/OE.420989.

\bibitem{Huang2022}
X.~Huang \textit{et al.}, Do-it-yourself radiative cooler as a radiative cooling standard and cooling component for device design. \textit{J.\ Photonics Energy} \textbf{2022}, \textit{12} (1), 012112. DOI: 10.1117/1.JPE.12.012112.

\end{thebibliography}
\end{document}